\documentclass[a4paper]{article}

\usepackage{INTERSPEECH2020}
\usepackage{threeparttable}
\usepackage{romannum}
\usepackage{multirow}
\usepackage{graphicx}
\usepackage{tikz,pgfplots}
\usetikzlibrary{patterns,calc}
\usepackage{comment}
\usepackage{mathtools}
\usepackage{hyperref}

\title{Semi-supervised Learning for Multi-speaker Text-to-speech Synthesis Using Discrete Speech Representation}
\name{Tao Tu, Yuan-Jui Chen, Alexander H. Liu, Hung-yi Lee}
\address{
  College of Electrical Engineering and Computer Science, National Taiwan University}
\email{\{r07922022, r07922070, r07922013, hungyilee\}@ntu.edu.tw}

\begin{document}

\maketitle
\begin{abstract}
Recently, end-to-end multi-speaker text-to-speech (TTS) systems gain success in the situation where a lot of high-quality speech plus their corresponding transcriptions are available.
However, laborious paired data collection processes prevent many institutes from building multi-speaker TTS systems of great performance.
In this work, we propose a semi-supervised learning approach for multi-speaker TTS.
A multi-speaker TTS model can learn from the untranscribed audio via the proposed encoder-decoder framework with discrete speech representation.
The experiment results demonstrate that with only an hour of paired speech data, whether the paired data is from multiple speakers or a single speaker, the proposed model can generate intelligible speech in different voices.
We found the model can benefit from the proposed semi-supervised learning approach even when part of the unpaired speech data is noisy.
In addition, our analysis reveals that different speaker characteristics of the paired data have an impact on the effectiveness of semi-supervised TTS.
\end{abstract}
\vspace{-2pt}
\section{Introduction}
% Praise TTS systems
% Problems
% semi-tts
Recent advances in the neural-based end-to-end text-to-speech (TTS) systems have closed the gaps between the human speech and synthesized speech in the aspects of both speech quality and speech intelligibility~\cite{oord2016wavenet, ping2018clarinet}. % cite: tacotron, wavenet, etc.
The notable results are shown not only for single speaker TTS modeling~\cite{wang2017tacotron, shen2018natural, taigman2017voiceloop, sotelo2017char2wav, ren2019fastspeech}, multi-speaker TTS modeling~\cite{ping2017deep, park2019multi} but also for cloning prosody style~\cite{wan2019chive, 49062, klimkov2019fine, lee2019robust, skerry2018towards} and speaker characteristics~\cite{wang2018style, hsu2018hierarchical,jia2018transfer, zhang2019learning}.
However, these achievements require large amounts of paired high-quality speech and text data (i.e. paired data), which is typically unavailable under the low-resource condition due to the laborious and expensive data collection and human labeling. 
Contrarily, unannotated speech data (i.e. unpaired data) is relatively prevalent and accessible.
Therefore, semi-supervised training of TTS that incorporates unpaired speech is crucial and worth investigating as it reduces the required amount of paired data for building a TTS system of high performance.

% semi-supervised way
% semi-supervised for multi-speaker
Semi-supervised learning for TTS has shown remarkable results in single speaker synthesis~\cite{ren2019almost, chung2019semi, liu2019towards}, where unpaired text or speech data are utilized to help the model training. 
Ren et al.~\cite{ren2019almost} proposed to jointly train a phoneme recognition model and a speech synthesis model with unpaired data. 
Chung et al.~\cite{chung2019semi} performed semi-supervised training on Tacotron~\cite{wang2017tacotron} in a pretrain-finetune manner. 
Different from the pretrain-finetune method, Liu and Tu et al.~\cite{liu2019towards} utilized the unpaired data for TTS training in an end-to-end manner.
They proposed Sequential Representation Quantization AutoEncoder (SeqRQ-AE) to learn discrete speech representation from a large amount of unpaired speech data.
With the aid from a few paired data, the discrete representations could be mapped to phonemes, and the model can be used for text-to-speech synthesis.

% Why this paper
Even though many efforts have been made to semi-supervised learning for TTS, prior works~\cite{ren2019almost, chung2019semi, liu2019towards} focused on single speaker TTS modeling and left multi-speaker TTS unstudied. 
Moreover, previous works~\cite{ren2019almost, liu2019towards} leverage only a large amount of unpaired speech from a single speaker which is also challenging to collect in practice.

% The proposed method
In this work, we move further to exploit semi-supervised multi-speaker TTS that can utilize unpaired speech with the supervision of only a few paired data (1 hour in total from either single speaker or multiple speakers).
We propose an extended architecture of SeqRQ-AE for semi-supervised multi-speaker TTS, where our framework consists of a phonetic encoder as in SeqRQ-AE, an extended speaker representation table, and a multi-speaker TTS model as the decoder.
The phonetic encoder transforms an utterance into a sequence of discrete phonetic representations by representation discretization and discrete representation mapping as in SeqRQ-AE.
The speaker representation table contains speaker representation for each speaker in the training set. 
The decoder takes the phonetic representations along with the speaker representation and decodes the corresponding speech.
This encoder-decoder framework can be jointly learned from unpaired data by imposing a reconstruction loss. 
Samples drawn from our model are provided on \url{https://ttaoretw.github.io/multispkr-semi-tts/demo.html}.

The contributions of this paper are highlighted as follows:
\begin{itemize}
    \item To the best of our knowledge, this is the first study of semi-supervised multi-speaker TTS.
    \item Our semi-supervised method matches the performance of the fully-supervised topline when only 1 hour of multi-speaker training data is annotated.
    \item When only 1 hour of single-speaker training data is labeled, our method can still generate intelligible speech of different voices.
    \item The effectiveness of semi-supervised multi-speaker TTS is further verified by considering the experiment with noisy unpaired speech data, which makes our method more feasible in practice.
    \item We take a closer look at the impact of speaker characteristics on the effectiveness of semi-supervised TTS.
\end{itemize}

% Overview picture

\begin{figure*}[ht]
\centerline{\includegraphics[width=0.9\linewidth]{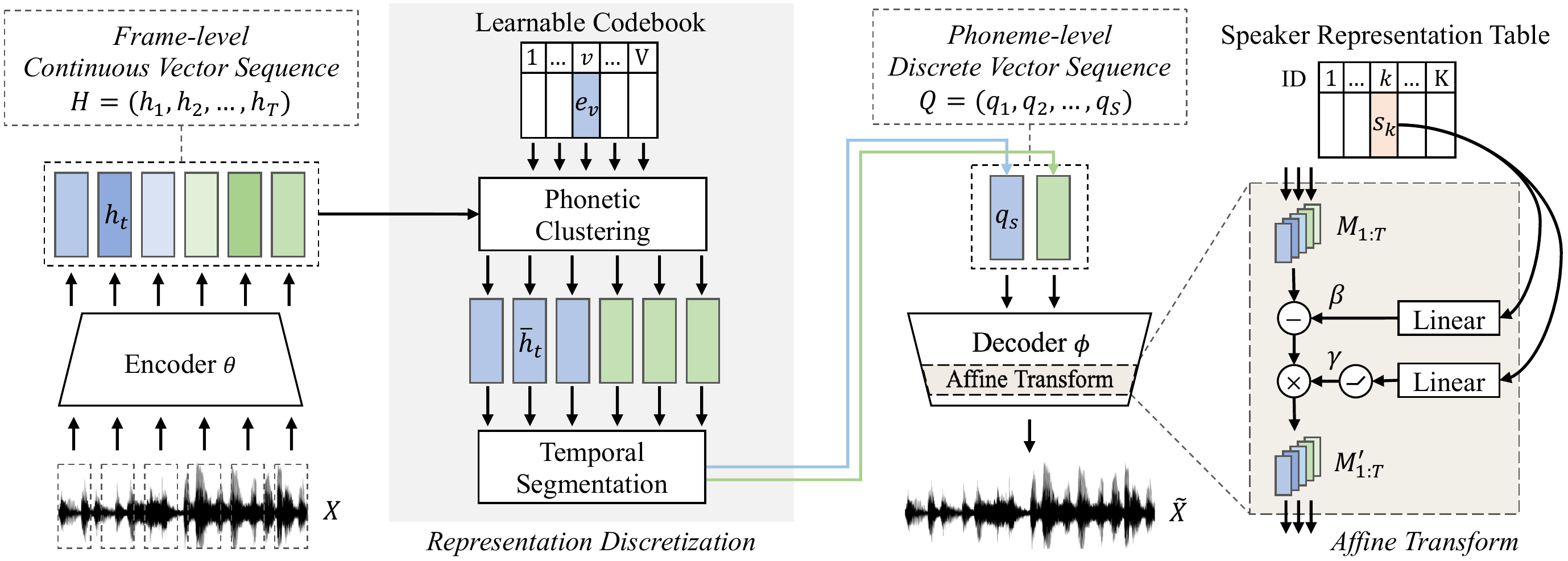}}

\caption{\small 
Overview of the proposed model.
The input speech $X$ is first encoded into the frame-level continuous vector sequence $H$.
Next, Representation Discretization (see Sec.~\ref{subsubsec:phn-enc}) is performed to obtain the phoneme-level discrete vector sequence $Q$.
$Q$ would be fed into a sequence-to-sequence decoder conditioned on speaker representation to reconstruct the input speech (see Sec.~\ref{subsec:multi-spkr-synth}).
}

\vspace{-8pt}
\label{fig:overview}
\end{figure*}
%\section{Proposed Method}
% The design idea of the framework
% The information in speech signals can be roughly categorized into two parts: the static part and the phonetic part.
% The static part stands for the information independent of time and merely changes within one utterance, which we see as the speaker characteristics.
% On the other hand, the phonetic part describes the information smoothly varying across time.
% In our proposed method, a phonetic encoder is used for extracting the phonetic information from speech in the form of discrete speech representation, and a speaker representation table is maintained for modeling static information.
% In addition, a decoder takes the discrete speech representations along with a speaker representation to generate speech.

% We organize the methodology as follows:
% In Section~\ref{subsec:phn-enc}, we explain the phonetic encoder with the representation discretization and discrete representation mapping mechanism.

% Introduce SeqRQ-AE
% \vspace{-6pt}
\section{Sequential Representation Quantization AutoEncoder (SeqRQ-AE)}
\label{subsec:SeqRQ-AE}
%We refer our readers to the prior work~\cite{liu2019towards} for more detailed description of such speaker-dependent framework.
In this section, we briefly overview the SeqRQ-AE, which is trained from a large amount of unpaired audio $X_\text{unpair}$ and limited audio-text pairs $(X_\text{pair},Y_\text{pair})$, where $Y_\text{pair}$ is the corresponding phoneme sequence of $X_\text{pair}$. 

\subsection{Phonetic Encoder}
\label{subsubsec:phn-enc}
% A speech signal may change dramatically during a sentence, however, its underlying phonetic information varies smoothly across time.
% Instead of modeling the speech signal directly, the phonetic encoder aims to capture the slow changing phonetic information.
Given a speech sequence $X = (x_1,x_2,...,x_T)$ of $T$ frames, an encoder network $\text{Enc}_{\theta}(\cdot)$ extracts the corresponding frame-level representation sequence
\begin{equation}
    \label{eq:enc}
    H \equiv (h_1,h_2,...,h_T) = \text{Enc}_\theta(X).
\end{equation}
To obtain the phoneme-level discrete speech representation sequence $Q=(q_1,q_2,...,q_S)$ that matches the underlying phoneme sequence, \textit{representation discretization} and \textit{discrete representation mapping} are applied.

% Phonetic clustering
\vspace{2pt}
% \noindent \textbf{Phonetic Clustering.}
\noindent \textbf{Representation Discretization.}
To perform representation discretization, a learnable \textit{codebook} $E=(e_1,e_2,...,e_V)$ of size $V$ is maintained, where each $e_i \in \mathbb{R}^D$ is called a \textit{codeword}.
For an encoded frame-level representation sequence $H$, the closest codeword $e_v$ is used as a substitute for each representation $h_t$, and this operation is called \textit{phonetic clustering}~\cite{liu2019towards}.
The gradient of this non-differentiable operation is approximated by straight-through (ST) gradient estimator~\cite{bengio2013estimating}.
The phonetic clustering process produces a codeword sequence $\bar{H}=(\bar{h}_1,\bar{h}_2,...,\bar{h}_T)$ of length $T$ where each element is one of the $V$ codewords.
Besides, to match the sequence length of the codeword sequence to the underlying phoneme sequence, \textit{temporal segmentation} is performed to group repeated consecutive codewords into one codeword. 

\begin{table*}[t]
% \small
% \footnotesize
\centering
\resizebox{\linewidth}{!}{
\begin{threeparttable}

\caption{Performance comparison of different methods. The subscript "(n)" indicates the MUSAN noises are added to the speech data. The naturalness MOS (\romannum{1}) is reported with 95\% confidence intervals. The recognition result (\romannum{2}) is reported with character error rate (CER).}
\begin{tabular}{ l| l| c| c| c| c| c}
\toprule
\multirow{2}{*}{\textbf{Experiment}}& \multirow{2}{*}{\textbf{Method}} & \multirow{2}{*}{\textbf{Paired Data}} & \textbf{Multi-speaker} & \multirow{2}{*}{\textbf{Unpaired Data}} & \textbf{(\romannum{1})} & \textbf{(\romannum{2})} \\
& & & \textbf{Supervised} & &  \textbf{Naturalness} & \textbf{CER} 
\\ \hline \\[-1em]
 \multirow{3}{*}{Baseline} & (b-1) Ground Truth & - & - & - & 4.88 $\pm$ 0.033 & 7.98 \\ \\[-1em]
 & (b-2) Tacotron-2 & VCTK-25\textsubscript{hr} & \checkmark (108\textsubscript{spkr}) & - & 3.59 $\pm$ 0.066 & 8.11 \\ \\[-1em]
 & (b-3) Tacotron-2 & VCTK-1\textsubscript{hr} & \checkmark (108\textsubscript{spkr}) & - & 1.47 $\pm$ 0.055 & 72.67 \\ \\[-1em]
 \hline\hline  \\[-1em]
  \multirow{4}{*}{Semi-supervised} & (s-1) Sp-chain~\cite{tjandra2017listening}\textsuperscript{\textdagger} & VCTK-1\textsubscript{hr} & \checkmark (50\textsubscript{spkr}) & VCTK-25\textsubscript{hr}-108\textsubscript{spkr} + LJ-other & 2.81 $\pm$ 0.071 & 31.30 \\ \\[-1em]
 & (s-2) Ours & VCTK-1\textsubscript{hr} & \checkmark (50\textsubscript{spkr}) & VCTK-25\textsubscript{hr}-108\textsubscript{spkr} + LJ-other & 3.46 $\pm$ 0.066 & 11.53 \\ \\[-1em]
 & (s-3) Sp-chain & LJ-1\textsubscript{hr} & - & VCTK-25\textsubscript{hr}-108\textsubscript{spkr} + LJ-other & 2.10 $\pm$ 0.065 & 45.47 \\ \\[-1em]
 & (s-4) Ours & LJ-1\textsubscript{hr} & - & VCTK-25\textsubscript{hr}-108\textsubscript{spkr} + LJ-other & 3.09 $\pm$ 0.073 & 21.70 \\ \\[-1em]
\hline \\[-1em]
 \multirow{4}{*}{~~~~~~-~w/ Noise} & (n-1) Ours & VCTK-1\textsubscript{hr} & \checkmark (50\textsubscript{spkr}) & VCTK-14\textsubscript{hr}-60\textsubscript{spkr} + LJ-other & 2.02 $\pm$ 0.087 & 41.95 \\ \\[-1em]
 & (n-2) Ours & VCTK-1\textsubscript{hr} & \checkmark (50\textsubscript{spkr}) & VCTK-14\textsubscript{hr}-60\textsubscript{spkr} + VCTK\textsubscript{(n)}-11\textsubscript{hr}-48\textsubscript{spkr} + LJ-other & 3.28 $\pm$ 0.073 & 12.78 \\ \\[-1em]
 & (n-3) Ours & LJ-1\textsubscript{hr} & - & VCTK-14\textsubscript{hr}-60\textsubscript{spkr} + LJ-other & 1.61 $\pm$ 0.069 & 80.07 \\ \\[-1em]
 & (n-4) Ours & LJ-1\textsubscript{hr} & - & VCTK-14\textsubscript{hr}-60\textsubscript{spkr} + VCTK\textsubscript{(n)}-11\textsubscript{hr}-48\textsubscript{spkr} + LJ-other & 2.85 $\pm$ 0.070 & 21.85 \\ \\[-1em]
\hline \\[-1em]
  \multirow{2}{*}{~~~~~~-~w/ different} & (c-1) Ours & LJ-1\textsubscript{hr} & - & VCTK-25\textsubscript{hr}-108\textsubscript{spkr} + LJ-other + MLJ-other & 3.22 $\pm$ 0.070 & 16.47 \\ \\[-1em]
 \multirow{2}{*}{~~~~~~~~Characteristics} & (c-2) Ours & MLJ-1\textsubscript{hr} & - &  VCTK-25\textsubscript{hr}-108\textsubscript{spkr} + LJ-other + MLJ-other & 2.31 $\pm$ 0.062 & 15.36 \\ \\[-1em]
 & (c-3) Ours & FLJ-1\textsubscript{hr} & - & VCTK-25\textsubscript{hr}-108\textsubscript{spkr} + FLJ-other + MLJ-other & 3.07 $\pm$ 0.079 & 16.20 \\ 
\bottomrule
\end{tabular}

\vspace{-2pt}
\begin{tablenotes}
\item{\textdagger} \small{\text{Trained without text-to-text cycle.}}
\end{tablenotes}
\vspace{-15pt}
\label{table:multispkr-tts}
\end{threeparttable}
}
\end{table*}

\vspace{2pt}
\noindent \textbf{Discrete Representation Mapping.}
To force each code of the codebook to be a phoneme, we first set the codebook size $V$ to be the number of all phonemes and assign each entry $e_v$ a phoneme $v$.
The paired speech data ($X_\text{pair}$, $Y_\text{pair}$) is used for learning the mapping. 
The probability of a continuous representation $h_t$ being mapped to a codeword $e_v$ is defined as
\begin{equation}
    \label{eq:phn_prob}
    P(v|h_t) = \frac{\exp(- \| h_t - e_v\|_2)}{\sum_{k \in V} \exp(- \| h_t - e_k\|_2)},
\end{equation}
and the probability for a frame-level phoneme sequence 
$\tilde{Y}=(v_1,v_2,...,v_T)$ can be approximated by
\begin{equation}
    \label{eq:seq_prob}
    P(\tilde{Y}|H)  \approx  \prod_{t=1}^{T}{P(v_t|h_t)}.
\end{equation}
Then, the connectionist temporal classification~\cite{graves2006connectionist} (CTC) is applied on the paired data with Eq.~(\ref{eq:seq_prob}) to maximize the log-likelihood of outputting target $Y_\text{pair}$.

\subsection{Speech Synthesizer}
\label{subsubsec:spkr-dependent-synth}
To reconstruct the input utterance, a decoder network $\text{Dec}_{\phi}(\cdot)$ takes the sequence of discrete speech representations $Q$ as input and synthesize audio $\widetilde{X}$ as below.
\begin{equation}
\begin{aligned}
    \label{eq:spkr-dependent-dec}
    \widetilde{X} = \text{Dec}_\phi(Q).
\end{aligned}
\end{equation}
In addition, the decoder can also do text-to-speech transformation by inputting code sequence $Q_\text{pair}$ retrieved from the codebook according to the ground truth phoneme sequence $Y_\text{pair}$.
The overall loss function can be written as
\begin{equation}
\label{eq:loss_func}
\begin{split}
L_\text{total} =
    &~\lambda\cdot\text{MSE}( \widetilde{X},X_\text{unpair}) \\
    &- \log P(Y_\text{pair}|H)\\
    &+ \text{MSE}(\text{Dec}_{\phi}(Q_\text{pair}),X_\text{pair}),
\end{split}
\end{equation}
where the first term is the reconstruction loss of unpaired speech $X_\text{unpair}$, the second term is the CTC loss for $Y_\text{pair}$, the last term is the TTS loss for target audio $X_\text{pair}$, and $\lambda$ is fixed to be 10 throughout the end-to-end training process.
For more details, please refer to the prior work~\cite{liu2019towards}.

\vspace{-3pt}
\section{Multi-speaker SeqRQ-AE}
\label{subsec:multi-spkr-synth}
In the previous work~\cite{liu2019towards}, both $X_\text{unpair}$ and $X_\text{pair}$ are produced by the same speaker.
Here we assume the audio is from multiple speakers\footnote{We assume the speaker identities of both the paired and unpaired audio data are known.}, and we extend the decoder in Sec.~\ref{subsubsec:spkr-dependent-synth} into a multi-speaker synthesizer. 

In order to perform multi-speaker synthesis (as shown in the right-hand side of Figure~\ref{fig:overview}), the decoding process in Eq.~(\ref{eq:spkr-dependent-dec}) is equipped with a learnable speaker representation table $\{s_1,...,s_k,...,s_K\}$, where each vector $s_k$ is the embedding of a speaker, and $K$ is the total number of speakers in $X_\text{pair}$ and $X_\text{unpair}$. 
With speaker representations, Eq.~(\ref{eq:spkr-dependent-dec}) is modified as below:
\begin{equation}
\begin{aligned}
    \label{eq:multi-spkr-dec}
    \widetilde{X} = \text{Dec}_\phi(Q, s_k),
\end{aligned}
\end{equation}
where the decoder is conditioned on the speaker representation $s_k$ obtained from the speaker representation table according to speaker identity of the input utterance.
The loss function to be optimized is the same as Eq.~(\ref{eq:loss_func}), except that $\text{Dec}_\phi(Q)$ is replaced with $\text{Dec}_\phi(Q, s_k)$.

To perform speaker adaptive synthesis, we proposed to modify the intermediate state of the decoder with an affine transformation.
The scaling factor $\gamma$ and the shifting magnitude $\beta$ for some particular speaker $k$ can be derived by
\begin{equation}
\begin{aligned}
    \label{eq:scale-shift-term}
    \gamma &= \text{ReLU}(W_{\gamma}s_k + b_{\gamma}),\\
    \beta &= W_{\beta}s_k + b_{\beta},
\end{aligned}
\end{equation}
where $s_k$ is the speaker representation of speaker $k$ and both $W$, $b$ are learnable parameters of linear projection layer.
With the scaling factor $\gamma$ and the shifting magnitude $\beta$, the affine transformation is performed on the intermediate state $M_t$ of the decoder at each timestep $t$ of the synthesis process
\begin{equation}
\begin{aligned}
    \label{eq:static-op}
    {M}^{'}_{t} &= \gamma (M_t - \beta).
\end{aligned}
\end{equation}
In practice, we employ Tacotron-2~\cite{shen2018natural} as the decoder $\text{Dec}_{\phi}(\cdot)$ of our framework, where Tacotron-2 itself contains an encoder (Taco-encoder) and a decoder (Taco-decoder).
Taco-decoder consists of 2 LSTM layers and 5 convolution layers, where we select the output hidden states of the first LSTM as the input of affine transformation $M_t$.
Afterward, the modified LSTM output $M^{'}_{t}$ is passed to the next layer.
We found that this affine transformation scheme makes the training of the multi-speaker TTS model more stable.

\vspace{-6pt}
\section{Experiment}
\subsection{Experiment Setup}
\label{subsec:setup}
\vspace{1pt}
\noindent \textbf{Model Architecture.}
For the phonetic encoder and the codebook, we follow the setup as in the prior work~\cite{liu2019towards}.
The Griffin-Lim algorithm~\cite{griffin1984signal} is applied to estimate the phases and converts spectrograms to waveforms as in Tacotron~\cite{wang2017tacotron}.
The differential spectral loss~\cite{shechtman2019sequence} is also adopted to boost the performance of the TTS model. 

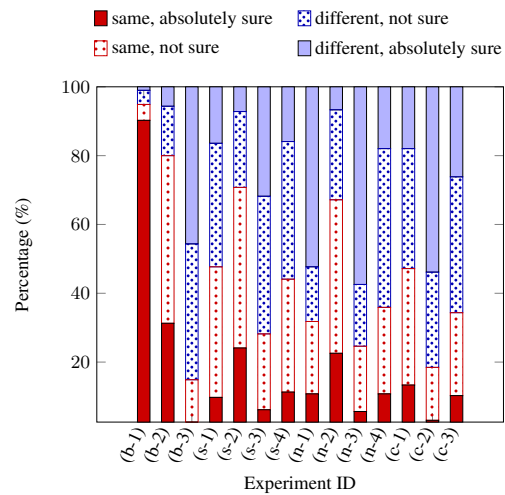
\begin{figure}
% \centering
\resizebox{.85\linewidth}{!}{
\begin{tikzpicture}
    \begin{axis}[
            ybar stacked,
            bar width=6pt,
            % nodes near coords,
            enlarge y limits=0,
            enlarge x limits=0.15,
            legend style={
                    draw=none,
                    at={(0.5, +1.27)},
                    legend cell align=left,
                    anchor=north,
                    legend columns=2,
                    inner sep=5pt,
                    /tikz/every even column/.append style={column sep=10pt}
                },
            transpose legend,
            xlabel style={
            at={(0.5, -0.05)}
            },
            xlabel={Experiment ID},
            ylabel={Percentage (\%)},
            symbolic x coords={
                    (b-1),
                    (b-2),
                    (b-3),
                    (s-1),
                    (s-2),
                    (s-3),
                    (s-4),
                    (n-1),
                    (n-2),
                    (n-3),
                    (n-4),
                    (c-1),
                    (c-2),
                    (c-3),
                },
            xtick=data,
            x tick label style={
                    rotate=65,
                    anchor=east,
                    align=left
                },
        ]
        \addplot[
            ybar,
            fill=red!80!black,
            % pattern=north east lines,
            % pattern color=red!80!black,
        ] plot coordinates {
                ((b-1), 90.26)
                ((b-2), 31.28)
                ((b-3), 2.56)
                ((s-1), 9.74)
                ((s-2), 24.10)
                ((s-3), 6.15)
                ((s-4), 11.28)
                ((n-1), 10.77)
                ((n-2), 22.56)
                ((n-3), 5.64)
                ((n-4), 10.77)
                ((c-1), 13.33)
                ((c-2), 3.08)
                ((c-3), 10.26)
            };
        \addplot[
            ybar,
            red!80!black,
            pattern=dots,
            pattern color=red!80!black,
        ] plot coordinates {
                ((b-1), 4.62)
                ((b-2), 48.72)
                ((b-3), 12.31)
                ((s-1), 37.95)
                ((s-2), 46.67)
                ((s-3), 22.05)
                ((s-4), 32.82)
                ((n-1), 21.03)
                ((n-2), 44.62)
                ((n-3), 18.97)
                ((n-4), 25.13)
                ((c-1), 33.85)
                ((c-2), 15.38)
                ((c-3), 24.10)
            };
        \addplot[
            ybar,
            blue!70!black,
            pattern=crosshatch dots,
            pattern color=blue!70!black,
        ] plot coordinates {
                ((b-1), 4.10)
                ((b-2), 14.36)
                ((b-3), 39.49)
                ((s-1), 35.90)
                ((s-2), 22.05)
                ((s-3), 40.00)
                ((s-4), 40.00)
                ((n-1), 15.90)
                ((n-2), 26.15)
                ((n-3), 17.95)
                ((n-4), 46.15)
                ((c-1), 34.87)
                ((c-2), 27.69)
                ((c-3), 39.49)
            };
        \addplot[
            ybar,
            fill=blue!30!white,
            % pattern=crosshatch,
            % pattern color=blue!80!black,
        ] plot coordinates {
                ((b-1), 1.03)
                ((b-2), 5.64)
                ((b-3), 45.64)
                ((s-1), 16.41)
                ((s-2), 7.18)
                ((s-3), 31.79)
                ((s-4), 15.90)
                ((n-1), 52.31)
                ((n-2), 6.67)
                ((n-3), 57.44)
                ((n-4), 17.95)
                ((c-1), 17.95)
                ((c-2), 53.85)
                ((c-3), 26.15)
            };
        \legend{
            \raisebox{3pt}{same, absolutely sure},
            \raisebox{3pt}{same, not sure},
            \raisebox{3pt}{different, not sure},
            \raisebox{3pt}{different, absolutely sure}
        }
    \end{axis}
\end{tikzpicture}
}
\vspace{-5pt}
\caption{\small The results of speaker similarity test. The x-axis labels indicate the experiment IDs as in Table~\ref{table:multispkr-tts}.
}
\label{fig:similarity}
\vspace{-15pt}
\end{figure}

\vspace{2pt}
\noindent \textbf{Datasets}
We use VCTK corpus~\cite{vctk}, which consists of read English speech data from 108 speakers with complete transcriptions.
After removing the leading and ending silence by Montreal Forced Aligner~\cite{mcauliffe2017montreal}, we have about 26 hours of speech data in total.
We randomly choose 1000 audio files for testing and other 1000 audio for selecting hyperparameters.
In addition, an hour data randomly chosen from LJSpeech~\cite{ljspeech17}, which is a 24 hours English dataset from a single female speaker, is used as the paired speech data (LJ-1\textsubscript{hr}) and the remaining data are used as unpaired speech data (LJ-other).
Moreover, we use Google cloud text-to-speech to synthesize MLJ and FLJ datasets based on the text from LJSpeech~\cite{ljspeech17}, where MLJ and FLJ are from a male and a female speaker, respectively.
Following the data partition of LJSpeech, MLJ and FLJ are also split into MLJ-1\textsubscript{hr}, FLJ-1\textsubscript{hr} and MLJ-other, FLJ-other.
We use \textit{x}\textsubscript{hr} to indicate the total amount of audio data (in hours), and \textit{x}\textsubscript{spkr} to indicate the number of used speakers where speech data size for each speaker is roughly equivalent.
The speakers in the test set will not appear in the paired training data set for all experiments except for the baseline experiments (b-2) (b-3).
As for text and audio preprocessing, we follow the prior work~\cite{liu2019towards}.

\vspace{2pt}
\noindent \textbf{Speech Naturalness Test.} 
The Mean Opinion Score (MOS) test is conducted for measuring speech quality, where 50 sentences are randomly chosen from the testing set and listened by 60 subjects.
The subjects are asked to rate the audio based on the speech naturalness.
The rating is on a 5-point scale in increments of 1.
The higher the MOS, the better the quality of the given audio.
Each audio file receives at least 6 ratings.
The results are shown in the col.~(\romannum{1}) of Table~\ref{table:multispkr-tts}.

\vspace{2pt}
\noindent \textbf{Content Correctness Test.} 
To analyze whether the model outputs correct speech content, we conduct Automatic Speech Recognition (ASR) test using the ASR service provided by Google Cloud Speech API to recognize synthesized audio or ground truth audio in the testing set.
Then, the character error rate (CER) is computed based on the ground truth texts.
The lower the CER, the more accurate the model output content is.
The results are shown in the col.~(\romannum{2}) of Table~\ref{table:multispkr-tts}.

\vspace{2pt}
\noindent \textbf{Speaker Similarity Test.} 
To measure the speaker similarity, the speaker similarity test~\cite{wester2016analysis} is conducted.
Given a pair of ground truth audio sample and TTS output sample from the same speaker with different contents, 60 subjects were asked to answer the question: "Do you think the same speaker has produced these two samples?" with options "same, absolutely sure", "same, not sure", "different, not sure" and "different, absolutely sure".
There are 50 randomly chosen pairs are answered by subjects, and each pair receives at least 6 answers.
The results are shown in Figure~\ref{fig:similarity}.

\vspace{2pt}
\noindent \textbf{Baseline.}
To objectively evaluate the effectiveness of our proposed method, we first perform the evaluation on raw audio from the test set and the original Tacotron-2 model with \textit{full supervision} to serve as our baseline as shown in Table~\ref{table:multispkr-tts}.
The Tacotron-2 model that is fully supervised by 25 hours (b-2) and 1 hour (b-3) of multi-speaker paired data can be viewed as the top-line and the bottom-line performance of our semi-supervised methods, respectively.

\vspace{-2pt}
\subsection{Multi-speaker Speech Synthesis}
\label{exp:multispkr-tts}
\vspace{1pt}
In this part ("Semi-supervised" partition of Table~\ref{table:multispkr-tts}), we also compared our method to the speech chain\footnote{Text-to-text cycle is not used since we found that the text-to-text cycle hurts the performance a lot when there is only one hour paired data available.}~\cite{tjandra2017listening} model (Sp-chain, row (s-1) and (s-3) in Table~\ref{table:multispkr-tts}) that shares the same architecture with our proposed model.
Sp-chain can be viewed as our model \textit{without} the learnable codebook and ST gradient estimator for unpaired speech data.

\vspace{2pt}
\noindent \textbf{Semi-supervised TTS w/ Multi-speaker Paired Data.} 
In this setting, 1 hour of paired data comes from 50 speakers (about 72 seconds for each speaker) are utilized for the TTS training.
% Naturalness
For speech naturalness (Table~\ref{table:multispkr-tts} col.~(\romannum{1})), content correctness (Table~\ref{table:multispkr-tts} col.~(\romannum{2})) , and speaker similarity  (Figure~\ref{fig:similarity}), we can see that the speech quality of our method (s-2) is consistently better than Sp-chain (s-1).
We conjecture this is because the proposed method allows the gradients to flow from the decoder through the encoder using the ST gradient estimator while Sp-chain does not.
This makes the encoder and the codebook also being updated to obtain superior representations for better speech reconstruction.
It is worth noticing that our method (s-2) \textit{matches the performance of the topline (b-2)} which has seen the speakers in the test set.
In the meanwhile, the bottom-line model (b-3) can hardly generate intelligible speech.
This demonstrates the effectiveness of semi-supervised TTS training.

\vspace{2pt}
\noindent \textbf{Semi-supervised TTS w/ Single Speaker Paired Data.}
% Single speaker setting
% What's more, we investigate the single speaker paired data setting, which is also possible in realistic.
In this setting, all paired data comes from a single speaker, which indicates that the model can only learn to synthesize different speaker from unpaired data.
Consistent with the setting of multi-speaker paired data, our method (s-4) outperforms the Sp-chain (s-3) in this setting.
By comparing (s-4) to (s-2), we find that the quality of synthesized speech deteriorates a bit when the paired data come from only one speaker.
Despite the performance drop compared to multi-speaker paired data setting, the multi-speaker TTS model trained in single speaker paired data setting is still much better than the baseline (b-3) in both naturalness and speaker similarity, which demonstrates the gain from semi-supervised learning.

% \vspace{-6pt}
\subsection{Impact of Noisy Unpaired Data}
\label{exp:noisy}
In this experiment, we discuss whether the proposed method can benefit from noisy unpaired data.
This is important because considerable high-quality clean audio is hard to collect and unpaired data are likely to be recorded in noisy environments in the real world case.
To simulate this situation, we take a part of VCTK data and manually add noise to it, which we refer to as $\text{VCTK}_{(n)}$ in the "w/ Noise" experiment of Table~\ref{table:multispkr-tts}.
The noises (10-30dB SNR) used in $\text{VCTK}_{(n)}$ are randomly selected from the MUSAN dataset~\cite{snyder2015musan}.
This synthetic noisy dataset includes 11 hours of speech from 48 speakers while the rest 14 hours of speech from 60 speakers in VCTK remains clean without noise.

Results with noisy unpaired data are reported in the "w/ Noise" experiment of Table~\ref{table:multispkr-tts}.
For the model (n-1) trained with 14 hours of clean unpaired data, the quality of the output speech is worse than the output of the model (n-2) trained with additional noisy unpaired data.
This can be seen from the speech naturalness, content correctness, and speaker similarity.
Besides, by comparing (n-2) with (s-2), we can see that the synthesis performance only drops a bit when some part of the unpaired data is noisy.
These demonstrate that the TTS model can benefit from unpaired data even if part of it is noisy.
The experiments in single speaker paired data setting are also conducted here (n-3) (n-4) and the same conclusion can be obtained.

% \vspace{-6pt}
% Discuss the weight of noisy cycle and clean cycle
\subsection{Impact of Speaker Characteristics of Paired Data}
\label{exp:spkr-char}
According to the result of the experiment (s-4), it is possible to construct a TTS model with only 1 hour paired data from a single speaker.
In this section, we would like to further study \textit{how the characteristics of the paired data influence the performance}. 
The results are reported in the ”w/ different Characteristics” experiment of Table 1.

First, we conduct (c-1) experiment with the paired data LJ-1\textsubscript{hr} from a female speaker and (c-2) experiment with the paired data MLJ-1\textsubscript{hr} from a synthesized male speaker\footnote{Synthesized audio is used here because we do not have large enough labeled clean audio from one male speaker.}.
We can see that the performance of (c-2) drops a lot no both in naturalness or in speaker similarity.%, which could be attributed to being male or synthesized.
To verify this drop comes from gender or  synthesized nature, we additionally change the data from LJ to FLJ which is from a synthesized female speaker (experiment (c-3)).
We observe that (c-3) deteriorates slightly than (c-1) with respect to naturalness and speaker similarity, which implies that training with synthesized speech only slightly hurt and the performance decline of (c-2) mainly comes from its male characteristic.
Therefore, we conclude that speaker characteristics are essential and female voice might be more applicable for semi-supervised learning when only single speaker paired data is available.

% \vspace{-8pt}
\section{Conclusion}
In this work, we study the semi-supervised multi-speaker TTS. 
Experiments show our proposed semi-supervised method matches the performance of the fully-supervised topline.
% We also found that the TTS model can benefit from a lot of unpaired speech data even if parts of them are noisy and the speaker characteristics of paired data influence the semi-supervised TTS performance.
In the future, we aim to explore the usage of the proposed method in cross-lingual settings~\cite{8683746, chen2019end}.

\bibliographystyle{IEEEtran}

\bibliography{main}

\end{document}